\documentclass[aps,twocolumn,amsmath,amssymb,10pt]{revtex4-1}
\usepackage{hyperref}
\usepackage{graphicx}
\usepackage{color}
\def\set#1{\mathbb{#1}}

\DeclareMathOperator{\trace}{tr} 

\def\ket#1{|#1\rangle}
\def\bra#1{\langle #1 |}
\def\bracket#1{\langle #1 \rangle}

\def\tenpow#1{^{\otimes #1}}
\def\cal#1{\mathcal #1}

\begin{document}
\title{Regularized maximum pure-state input-output fidelity of a quantum channel}
\author{Moritz F.\ Ernst and  Rochus Klesse}
\affiliation{Universit\"at zu K\"oln, Institut f\"ur Theoretische Physik,
  Z\"ulpicher Stra{\ss}e 77, 50937 K\"oln, Germany} \date{December 12,
2017}
\begin{abstract}
As a toy model for the capacity problem in quantum information theory
we investigate finite and asymptotic regularizations of the maximum
pure-state input-output fidelity $F(\cal N$) of a general quantum
channel $\cal N$. We show that the asymptotic regularization $\tilde
F(\cal N$) is lower bounded by the maximum output $\infty$-norm
$\nu_\infty(\cal N)$ of the channel. For $\cal N$ being a Pauli
channel we find that both quantities are equal. 
\end{abstract}

\maketitle

\section{Introduction}
An open problem of quantum information theory is finding an efficient
method to compute certain information capacities of a general quantum
channel $\cal N$, for instance, its capacity for transmission of
classical, private-classical or quantum information \cite{Wil13}. 
The problem arises because, according to the present state of
the theory, determining such a capacity $C(\cal N)$ requires regularizing a
corresponding single-shot capacity $C^{(1)}(\cal N)$ as
\begin{align}
C(\cal N) = \lim_{n\to\infty} \frac{1}{n} C^{(1)}(\cal N\tenpow n)\:.
\end{align}
The computation of $C^{(1)}(\cal N\tenpow n)$ involves typically maximization of an
entropic expression over the input quantum states of the $n$-times
replicated channel $\cal N\tenpow n$. This renders determining $\cal
C(\cal N)$ in general an analytically as well as numerically formidable problem, to which
presently no good solution is available. 

We do not attempt to solve any of the above capacity problems. Instead,
here we study as a toy problem a structurally similar but technically
far less demanding problem, namely regularizing the  maximum
pure-state input-output fidelity $F(\cal N)$ of a quantum channel
$\cal N$ \cite{gate-fidelity}. We find that in general the $n$-th
regularization $F^{(n)}(\cal N) = F(\cal N\tenpow n)^{1/n}$ 
shows a non-trivial $n$-dependence, as it is seen e.g.\ for
certain Pauli channels  (cf.\ Fig.\
\ref{fig1}, \ref{fig2}, or \ref{fig3}). Determining the asymptotic regularization
$\tilde F(\cal N) = \lim_{n\to\infty}F^{(n)}(\cal N)$ therefore represents a
problem that is structurally similar to determining capacities of a quantum channel. 
By employing symmetric trial states we show that the maximum output $\infty$-norm
$\nu_\infty(\cal N)$ is an easily-computable single-letter lower bound of the asymptotic
regularization $\tilde F(\cal N)$ for a general channel $\cal
N$. Moreover, from a result of King \cite{Kin02}, stating that the
maximum output $\infty$-norm is multiplicative for unital qubit
channels,  it follows that for a general Pauli channel 
$\nu_\infty$ actually coincides with $\tilde F$.  
This establishes $F$ on Pauli channels as a simple toy model
within which  non-trivial $n$-th regularizations are observed while at the same
time the asymptotic regularization $\tilde F$ is available.

\section{Notations, Definitions, Relations}\label{sec-notations}
We consider a quantum channel $\cal N$ with identical input and output
Hilbert space $\cal H$ of finite dimension $d$. I.e.\ $\cal N$ is 
a completely positive and trace preserving endomorphism on the linear operators on $\cal H$.
The Hermitian conjugate of the channel $\cal N$ with respect to the
Hilbert-Schmidt inner product $(A,B) = \trace A^\dagger B$ will be
denoted by $\cal N^\dagger$. For a pure-state (i.e. rank-1) density operator $\psi$ on $\cal H$
let 
\begin{align*}
F(\cal N, \psi) = \trace \psi \cal N(\psi)
\end{align*}
be the input-output fidelity of the channel $\cal N$ on $\psi$. As usual, we denote a state
vector of $\cal H$ and its dual by $\ket \psi$ and $\bra \psi$,
respectively, and the rank-1 density operator of the associated pure
state by $\psi$.

For the channel $\cal N$ we define the maximum input-output fidelity
$F(\cal N)$ on pure states, its $n$-th regularization $F^{(n)}(\cal N)$, and its
asymptotic regularization $\tilde F(\cal N)$ as
\begin{align*}
F(\cal N) \: &= \: \max_{\psi} \: F(\cal N,\psi)\:, \\
F^{(n)}(\cal N) &= \: F(\cal N \tenpow n)^{1/n}\:, \\
\tilde F(\cal N) \:  &= \: \lim_{n\to \infty} F^{(n)}(\cal N)\:,
\end{align*}
where the maximum is taken with respect to rank-1 density operators $\psi$. 
We will also need the maximum output $q$-norm $\nu_q(\cal N)$ of $\cal N$, its
$n$-th regularization $\nu_q^{(n)}(\cal N)$, and its asymptotic
regularization $\tilde \nu_q(\cal N)$, defined by  
\begin{align*}
%\:, \\
\nu_q(\cal N) \: &= \: \max_{\psi} || \cal N(\psi)||_q
\:, \\
\nu_q^{(n)}(\cal N) &= \: \nu_q(\cal N\tenpow n)^{1/n}\:, \\
\tilde \nu_q(\cal N) \: &= \: \lim_{n\to \infty} \nu_q^{(n)}(\cal N)\:.
\end{align*}
Note that $\nu_2(\cal N)$ can be expressed by the maximum input-output
fidelity as
\begin{align*}
\nu_2 (\cal N) = F(\cal N^\dagger \cal N)^{1/2}\:,
\end{align*}
since
\begin{align*}
|| \cal N(\psi)||^2_2 = (\cal N(\psi), \cal N(\psi)) = (\psi, \cal
N^\dagger \cal N(\psi )) = \trace \psi \: \cal N^\dagger\cal N(\psi)\:.
\end{align*} 
Another obvious relation is 
\begin{align}\label{upper_bound}
 F \: \le \: \nu_\infty \:, 
\end{align}
following from the fact that the maximum norm $||A||_\infty$ of an
operator $A$ can be expressed as $||A||_\infty = \max_\phi \trace \phi
A$, where the maximum is taken w.r.t.\ rank-1 density operators
$\phi$, and so
\begin{align*}
F(\cal N) \: & = \: \max_\psi \: \trace \psi \cal N(\psi) \\
& \: \le \max_\psi \: \max_ \phi \: \trace \phi \cal N(\psi) \\
& \: = \: \max_\psi \: || \cal N(\psi) ||_\infty = \nu_\infty(\cal N)\:.
\end{align*} 
Despite being an {\em upper} bound of the single-shot maximum fidelity, $\nu_\infty$
is also a {\em lower} bound of the regularized maximum fidelity,
\begin{align}\label{lower_bound}
\tilde F \: \ge \: \nu_\infty \:, 
\end{align}
as we  will be prove below in Sec.\ \ref{sec-lower_bound}. This
inequality cannot be an equality in general because  $\tilde F$ is
weakly multiplicative \cite{multiplicativity} whereas $\nu_\infty$ is
known to be not weakly multiplicative \cite{WH02}. 
Nevertheless, after regularization we obtain 
\begin{align}\label{regularized_lower_bound}
\tilde F \: \ge \: \tilde \nu_\infty \:,
\end{align}
which together with the regularized version of relation (\ref{upper_bound})
eventually proves
\begin{align}\label{regularizedF}
\tilde F\: = \: \tilde \nu_\infty\: 
\end{align}
(cf.\ Sec.\ \ref{sec-lower_bound}).
This equality of the regularized quantities does not
look very promising. However, with the aforementioned result of King
\cite{Kin02}, we can use it to compute the regularized
maximum fidelity for Pauli channels in Sec.\ \ref{sec-pauli}.

\section{$\tilde F \: \ge \:  \tilde\nu_\infty$}\label{sec-lower_bound}
First, we will prove 
\begin{align}\label{inequality}
\tilde F(\cal N) \: \ge \: \nu_\infty(\cal N)
\end{align}
for a fixed but arbitrary channel  $\cal N$. 
For this specific $\cal N$ let $\ket{\phi_1}, \ket{ \phi_2} \in \cal H$
such that $\nu_\infty(\cal N) = \bracket{\phi_1 | \cal N(\phi_2)| \phi_1}$.
It may happen that $\ket{ \phi_1} $ and $\ket{ \phi_2}$ are linearly
dependent. In this case $\nu_\infty(\cal N) = \bracket{\phi_1 | \cal N(\phi_1)| \phi_1}$
and the inequality (\ref{inequality}) holds trivially since 
\begin{align*}
\tilde F(\cal N) \ge F(\cal N\tenpow n, \phi_1\tenpow n)^{1/n} =
  \trace \phi_1 \cal N(\phi_1) = \nu_\infty(\cal N)
\end{align*}
for arbitrary $n$. For the rest of the proof we can therefore
assume without loss of generality that $\ket{ \phi_1}$ and $\ket{
  \phi_2}$ are linearly independent, and that 
\begin{align*}
\nu_\infty(\cal N) \: > \: \bracket{\psi | \cal N(\psi)| \psi}
\end{align*}
for all $\ket \psi \in \cal H$. Beyond that, we will also assume that 
$\bracket{\phi_1 | \cal N( \ket{\phi_2} \bra{\phi_1}) |\phi_2}$ is a
non-negative real number, which can be always achieved by multiplying
$\ket{\phi_1}$ with an appropriate phase.

Based on $\ket {\phi_1}$ and $\ket {\phi_2}$ let a sequence
$\ket{\psi_n}$ of state vectors be given by 
\begin{align*}
\ket{\psi_n} = \frac{c_n}{\sqrt 2} \left( \ket{\phi_1}\tenpow n \: + \:
  \ket{\phi_2}\tenpow n \right), \quad n \in \set N\:,
\end{align*}
where $c_n =\left( 1 + \Re \bracket{\phi_1|\phi_2}^n\right)^{-1/2}$ ensures normalization of $\ket{\psi_n}$.
Note that by assumption $|\bracket{\phi_1|\phi_2}|<1$ and hence
$\lim_n c_n = 1$. In the following we will show that 
\begin{align}\label{limit}
\lim_n F(\cal N\tenpow n, \psi_n)^{1/n} =
  \nu_\infty(\cal N), 
\end{align}
which suffices to prove the inequality (\ref{inequality}).
To do so, we start with expanding $F(\cal N\tenpow n, \psi_n)$ in
terms of $n$-th powers of coefficients 
\begin{align*}
\cal N_{ijkl} = \bracket{ \phi_i| \cal N( \ket{\phi_j} \bra{\phi_k})|
  \phi_l}, \quad i,j,k,l \in \{1,2\},
\end{align*}
as 
\begin{align*}
F(\cal N\tenpow n, \psi_n)= \frac{c_n^4}{4} \sum_{ijkl} ( \cal N_{ijkl})^n\:.
\end{align*}
Clearly, for large $n$ the sum is dominated by those coefficients
 $\cal N_{ijkl}$ which have maximal absolute value, namely, 
$\cal N_{1221}=\nu_\infty(\cal N)$ and potentially $\cal N_{2112}$, $\cal N_{1212}$, and
$\cal N_{2121}$. The sequence $F(\cal N\tenpow n,\psi_n)^{1/n}$ will
thus converge to $\cal \nu_\infty(\cal N)$, provided that there will be no accidental
cancellation among the terms. To show that this is actually the case,
we represent $\cal N$ with suitable Kraus operators $A_1, \dots, A_K$ as
\begin{align*}
\rho \: \mapsto \: \cal N(\rho) = \sum_{\nu = 1}^K A_\nu \rho
  A_\nu^\dagger\:, 
\end{align*}
and define four $K$-dimensional complex vectors
\begin{align*}
w \equiv r_{11}, \quad
x \equiv r_{12}, \quad
y \equiv r_{21}, \quad
z \equiv r_{22}
\end{align*}
by
\begin{align*}
(r_{ij})_\nu = \bracket{\phi_i| A_\nu | \phi_j}\:.
\end{align*}
This allows us to write
\begin{align*}
\cal N_{ijkl} = r_{lk}^\dagger r_{ij}
\end{align*} 
and hence
\begin{align}\label{sum}
F(\cal N\tenpow n, \psi_n) \: = \: \frac{c_n^4}{4}\sum_{ a,b \in
  \{w,x,y,z\}} (a^\dagger b)^n\:.
\end{align}
By the properties of $\ket{\phi_1}$ and $\ket{\phi_2}$ 
with respect to $\cal N$ and the Cauchy-Schwarz inequality we find that 
\begin{align*}
|x|^2 & = \cal N_{1221} = \nu_\infty(\cal N), \\
|y|^2 & = \cal N_{2112} \le |x|^2, \\
|w|^2 & = \cal N_{1111} < |x|^2, \\
|z|^2 & = \cal N_{2222} < |x|^2, \\
0 \le \cal N_{1212} &= y^\dagger x \le |y||x| \le |x|^2, \\
0 \le \cal N_{2121} &= x^\dagger y \le |x||y| \le |x|^2, \\
|a^\dagger b| < |x|^2 &\: \mbox{if $a$ or $b$ in $\{w,z\}$}.
\end{align*}
This means that all terms in $\{a^\dagger b\}_{a,b\in\{w,x,y,z\}}$ that are of maximal
absolute value $|x|^2=\nu_\infty(\cal N)$ are real positive numbers, which with
Eq. (\ref{sum}) immediately shows (\ref{limit}) and so concludes the
proof of the inequality (\ref{lower_bound}). 
 
The inequality (\ref{regularized_lower_bound}) follows then by
regularization: To this end we employ the weak multiplicativity of $F$
(\cite{multiplicativity}), by which 
\begin{align*}
\tilde F(\cal N) \: = \: \tilde F(\cal N\tenpow m)^{1/m} \: \ge \: \nu_\infty(\cal N\tenpow m)^{1/m}\:.
\end{align*}
This holds for all positive, integer $m$ and thus proves $\tilde F(\cal N) \ge
\tilde \nu_\infty(\cal N)$, which is inequaltiy
(\ref{regularized_lower_bound}). From inequality (\ref{upper_bound})
it is clear that also $\tilde F(\cal N) \le \tilde \nu_\infty(\cal
N)$, and therefore actually $\tilde F(\cal N) = \tilde \nu_\infty(\cal
N)$ for any channel $\cal N$, which shows Eq. (\ref{regularizedF}).

\section{Pauli Channel}\label{sec-pauli}
As an example we study the maximum input-output fidelity of a general
Pauli channel $\cal P$ on a qubit ($d = 2$), defined as 
\begin{align*}
\cal P(\rho) = \sum_{\alpha=0}^3 p_\alpha \: \sigma_\alpha \, \rho\,
  \sigma_\alpha\:, 
\end{align*}
where $\sigma_0$ is the identity, $\sigma_1,\sigma_2,\sigma_3$ are the
standard Pauli operators, and the non-negative coefficients $p_\alpha$
sum up to unity. Without loosing generality we demand that $p_1 \le
p_2 \le p_3$. 

It is not difficult to show that
\begin{align*}
\nu_\infty(\cal P) = 
\begin{cases}
p_0 + p_3 \: :\quad \mbox{for $p_0 \ge p_2\quad$ (a)} \\
p_2 + p_3 \: :\quad \mbox{for $p_0 < p_2\quad$ (b)} \\
\end{cases}.
\end{align*}
In the first case, (a), we find 
\begin{align*}
\nu_\infty(\cal P) = \bracket{\phi_3| \cal P(\phi_3) |\phi_3}\:, 
\end{align*}
where $\ket{\phi_3}$ is an eigenstate of $\sigma_3$ (either for
eigenvalue $+1$ or $-1$), while for the second case, (b), 
\begin{align*}
\nu_\infty(\cal P) = \bracket{\phi_+ | \cal P(\phi_-) | \phi_+}
\end{align*}
with $\ket{\phi_+}$ and $\ket{\phi_-}$ eigenstates of $\sigma_1$ for eigenvalues
$+1$ and $-1$, respectively. 

According to King \cite{Kin02}, the maximum $q$-norm $\nu_q$ of any
unital qubit channel $\Phi$ is multiplicative for any $q\ge 1$. This
means that for any other qubit channel $\Omega$
\begin{align*}
\nu_q(\Omega \otimes \Phi) = \nu_q(\Omega) \nu_q(\Phi)\:.
\end{align*}
Since $\cal P$ is unital this result particularly implies that its maximum
output $\infty$-norm does not change under regularization, i.e. 
for any $n$ 
\begin{align*}
\nu_\infty(\cal P) \: = \: \nu_\infty^{(n)}(\cal P) \:,
\end{align*}
and so clearly 
\begin{align*}
\nu_\infty(\cal P) \: = \: \tilde
  \nu_\infty(\cal P)\:.
\end{align*}
Thanks to this fortunate situation we can actually use our result Eq.\
(\ref{regularizedF}) to determine the asymptotic regularized
input-output fidelity of a Pauli channel as 
\begin{align}\label{barF}
\tilde F(\cal P) = 
\begin{cases}
p_0 + p_3 \: :\quad \mbox{for $p_0 \ge p_2\quad$ (a)}\\
p_2 + p_3 \: :\quad \mbox{for $p_0 < p_2\quad$ (b)}\\
\end{cases}.
\end{align}
In the first case, (a), $F(\cal P, \phi_3)=p_0+p_3 = F(\cal P \tenpow
n, \phi_3 \tenpow n)^{1/n}$, and thus regularization has no effect: $F(\cal P) =F^{(n)}(\cal P) =
\tilde F(\cal P)$.
In the second case, (b), we observe that 
\begin{align*}
F(\cal P) = F(\cal P, \phi_3) = p_0+p_3\:, 
\end{align*}
showing that here regularization increases the fidelity as  
\begin{align*}
F(\cal P) = p_0 + p_3 \: < \: p_2+p_3 = \tilde F(\cal P)\:.
\end{align*} 

\section{Numerical Results for Pauli Channels}\label{sec-numerical}

Taking the limit $n$ to infinity is essential in deriving the lower
bound $\nu_{\infty}$ of $\tilde F$. For this reason we do not have analytic results for the
$n$-th regularization for finite $n$. To obtain some insight into the
$n$-dependence of the regularized fidelity, we determined $F^{(n)}$ for
finite $n$ for three exemplary Pauli channels by numerical maximization with a
variant of the Barzilai-Borwein gradient method \cite{BB88}. 
Up to $n=6$ we maximized over the entire $n$-qubit Hilbert space.  Beyond
that, up to $n=26$, we restricted the maximization to symmetric state
vectors, i.e. to state vectors that do not change under permutation of
qubits. For all three investigated channels we found that for $n\le 6$
maximization over the symmetric state vectors and maximization over
all states gave identical values within numerical precision of $10^{-6}$.

\begin{figure}
\includegraphics[width=0.5\textwidth, page=1]{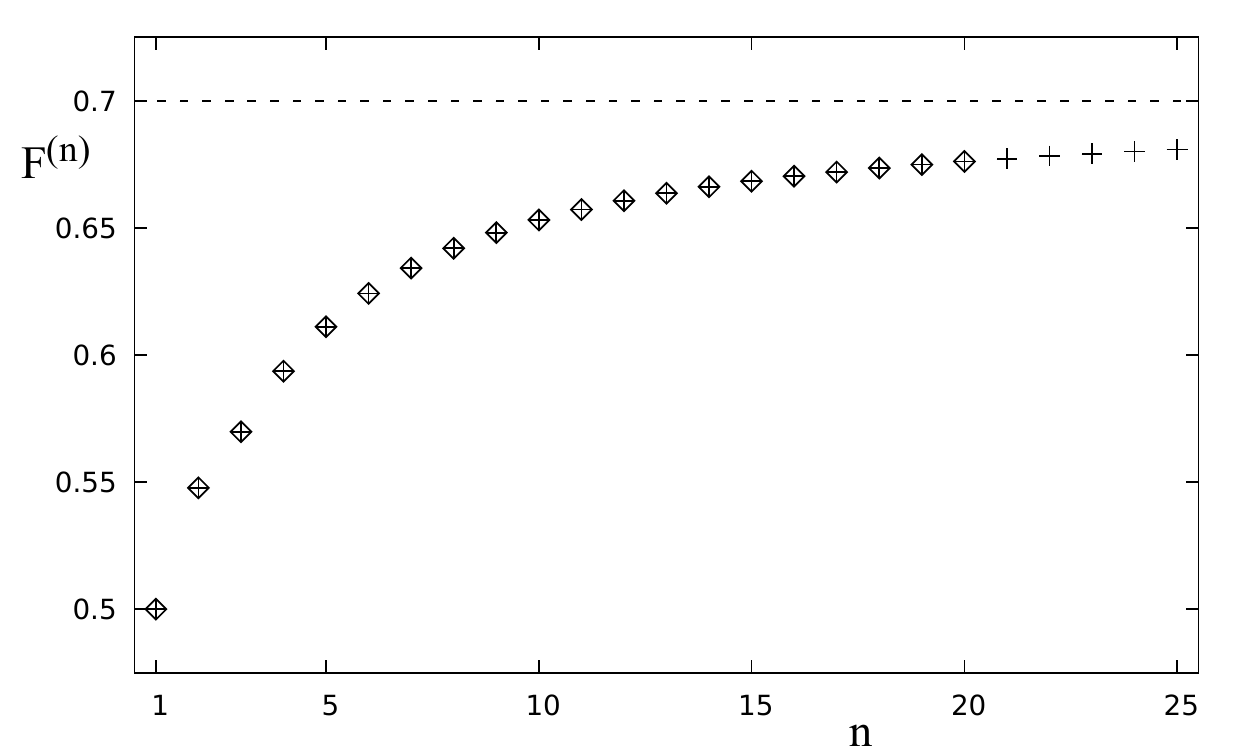}
 \caption{ \label{fig1} 
   Regularized maximum input-output fidelity for a Pauli channel $\cal P$
   with probabilities $p=(0.1,0.2,0.3,0.4)$. \\
   $\diamond$ :  $F^{(n)}(\cal P)$ determined by numerical maximization, \\
   $+$ : $F(\cal{P}\tenpow n, \psi_n)^{1/n} $ for trial states $\psi_n$,\\
   dashed line : $\tilde  F(\cal P)$.
}
\end{figure}

The first Pauli channel is given by probabilities
$p=(p_0,p_1,p_2,p_3)=(0.1,0.2,0.3,0.4)$. Since here $p_0 < p_2$,
according to Eq. (\ref{barF}) the asymptotic regularization of the
input-output fidelity is $\tilde F(\cal P)= 
p_2+p_3=0.7$, indicated by the dashed line in Fig.\ \ref{fig1}. 
The numerically determined maximum $n$-fidelities $F^{(n)}(\cal P)$ 
increase strictly monotonically  and evidently approach $\tilde F(\cal P)$ as $n$ increases 
from 1 to 20 (cf.\ $\diamond$-symbols in Fig.\ \ref{fig1}). 
These $n$-fidelities agree within numerical precision with 
$F(\cal{P}\tenpow n, \psi_n)^{1/n} $ on trial states
\begin{align}\label{psi_n}
\ket{\psi_n} = \frac{1}{\sqrt 2}( \ket{\phi_+}\tenpow n +
\ket{\phi_-}\tenpow n) \:,
\end{align}
which can be easily computed to be  
\begin{align}
F(\cal{P}\tenpow n, \psi_n)^{1/n} \: = \:
\frac{1}{2^{\frac{1}{n}}} &\left[  \:(p_0+p_1)^n +  \:
(p_0-p_1)^n \: + \right. \nonumber \\
& \: \left. (p_3-p_2)^n +  \:
(p_3+p_2)^n  \: \:
\right]^{\frac{1}{n}}\:\label{fidelity_trial}
\end{align} 
(cf.\ $+$-symbols in Fig.\ \ref{fig1}). We emphasize that this agreement is
coincidental, since we only proved
$\tilde F(\cal P) = \lim_n F(\cal P\tenpow n, \psi_n)^{1/n}$
(cf. Eq. (\ref{limit})).

\begin{figure}
\includegraphics[width=0.5\textwidth, page=1]{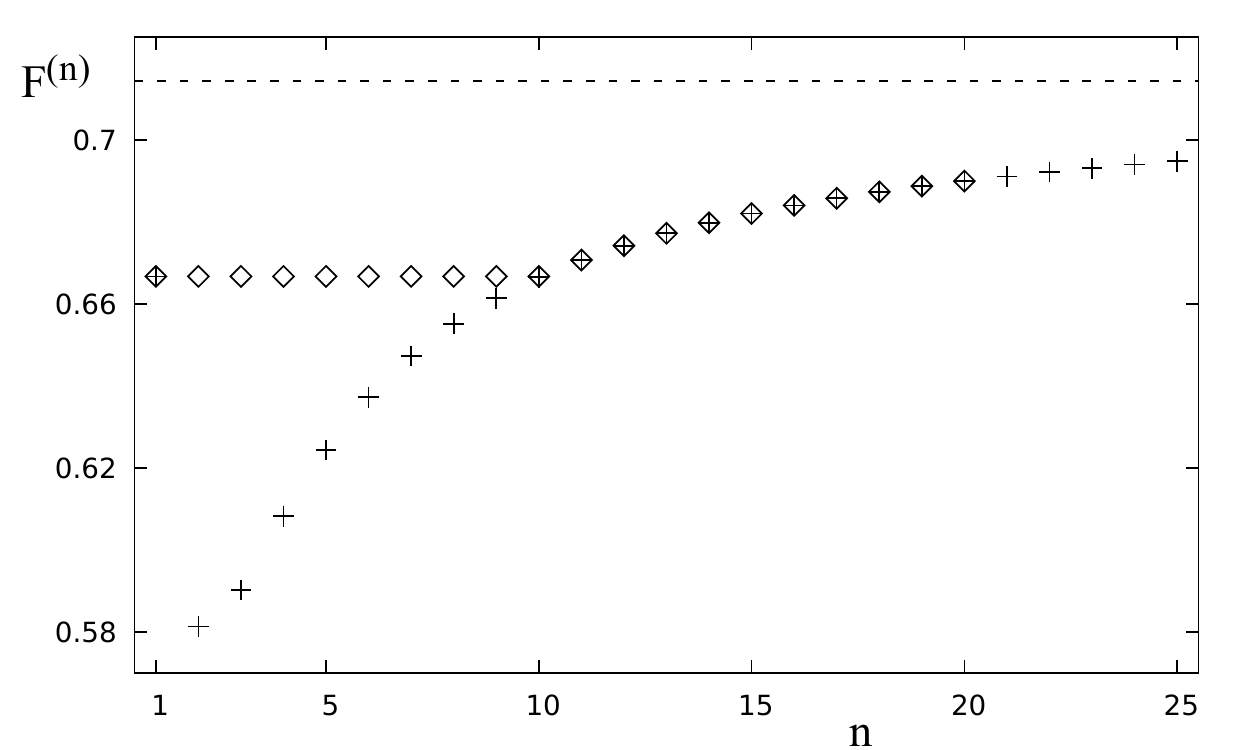}
 \caption{ \label{fig2} 
   Regularized maximum input-output fidelity for a Pauli channel
   with probabilities $p=(0.286, 0, 0.333, 0.381)$. Meaning of symbols as in Fig.\ 1.
}
\end{figure}

We studied a second Pauli channel with probabilities
\begin{align}\label{eps_prob}
p=(\frac{1}{3}- \epsilon, 0, \frac{1}{3}, \frac{1}{3}+\epsilon)
\end{align}
where $\epsilon = 1/21 = 0.04762$. As shown in Fig.\ \ref{fig2}, here
the numerically determined fidelities $F^{(n)}$ are constant of value
$F^{(1)}$ for $n\le 10$ and increase only for larger $n$ in order to 
approach asymptotically $\tilde F(\cal P)= p_2+p_3=\frac{2}{3}
+ \epsilon= 0.714$. Fittingly, the regularized fidelities $F(\cal 
P, \psi_n)^{1/n}$ for the trial states Eq.\ (\ref{psi_n}) are
submaximal for $1 < n < 10$, but appear again 
to be maximal for $n \ge 10$.
\begin{figure}
\includegraphics[width=0.5\textwidth, page=1]{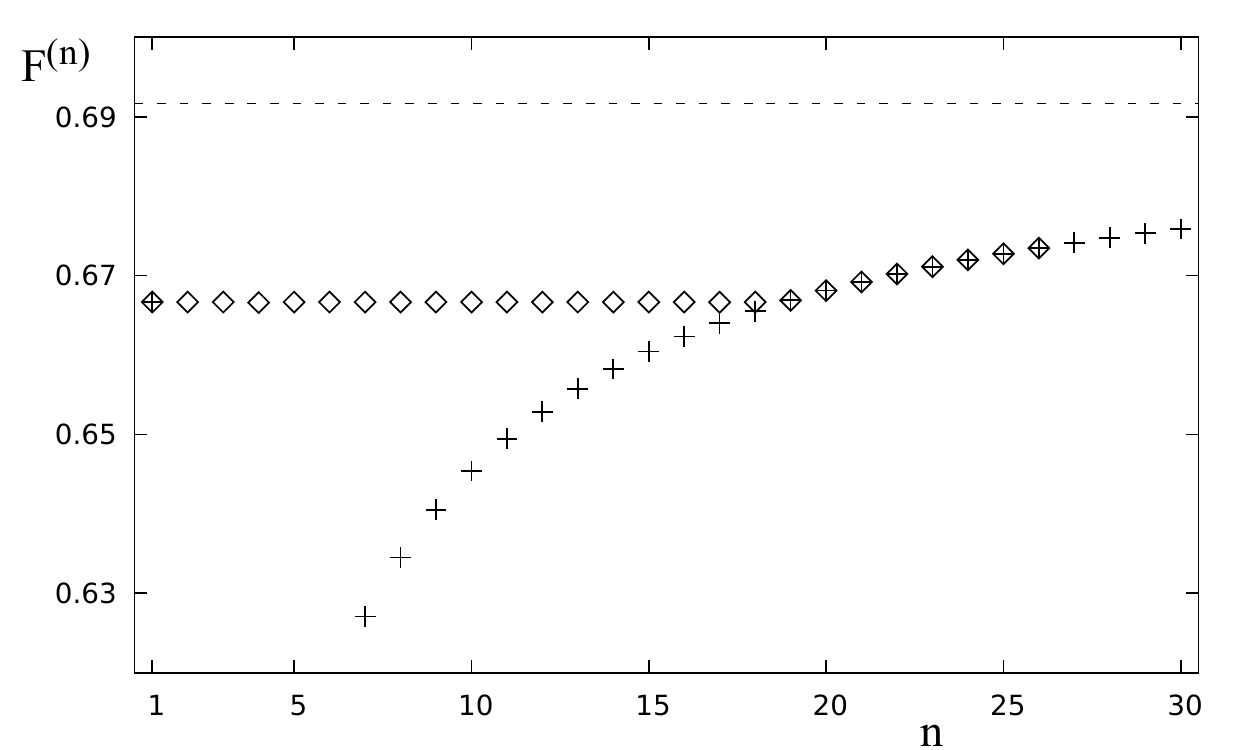}
 \caption{ \label{fig3} 
   Regularized maximum input-output fidelity for a Pauli channel
   with probabilities $p= (0.308, 0, 0.333, 0.358)$. Meaning of symbols as in Fig.\ 1.
}
\end{figure}

Closer inspection of Eq.\ (\ref{fidelity_trial}) 
reveals that for sufficiently small $\epsilon$ the fidelity
$F(\cal P\tenpow n,\psi_n)^{1/n}$
exceeds $F^{(1)}(\cal P)$  at 
\begin{align}
n \approx n_0 = \frac{\ln4}{3} \: \frac{1}{\epsilon}\:.
\end{align} 
If we take it for granted that the behavior shown in Fig. \ref{fig2}
is representative for sufficiently small $\epsilon $ it is clear
that for arbitrarily large $n_0$ one can always find a Pauli channel $\cal P_0$ such that 
$F^{(n)}(\cal P_0)  = F^{(1)}(\cal P_0)$  for $ n \le n_0 $
while  $F^{(n)}(\cal P_0) > F^{(1)}(\cal P_0)$ for $ n \ge n_0 $.

This is confirmed by our last numerical example presented in Fig.\
\ref{fig3}. Here we investigated a Pauli channel again with
probabilities as in Eq.\ (\ref{eps_prob}), but with $\epsilon = 0.025$,
leading to $\lceil n_0 \rceil = 19$. 

\section{Summary}\label{sec-conclusion}
We addressed the problem of determining finite and asymptotic regularizations
of the maximum input-output fidelity of a general quantum
channel. 
Using symmetric trial states we showed that the maximum output
$\infty$-norm $\nu_\infty(\cal N)$ is a lower bound of the
asymptotically regularized maximum input-output fidelity $\tilde F(\cal N)$ for
a general quantum channel $\cal N$. Moreover, for $\cal N$ being a
Pauli channel we found that a result of King already implies equality of $\nu_\infty(\cal
E)$ and $\tilde F(\cal N)$. 
Numerically determined finite regularizations $F^{(n)}(\cal P)$ for 
Pauli channels show a non-trivial $n$-dependence and confirm the
results for the asymptotic regularization. 

Financial support from DFG Grant No. ZI-513/1-2, from the center for
Quantum Matter and Materials of the University
of Cologne, and from the SFB/TR 12 is gratefully acknowledged.


\begin{thebibliography}{10}

\def \pra#1#2#3#4{ #1, Phys.\ Rev.\ A {\bf #2}, #3 (#4)} 
\def \prb#1#2#3#4{ #1, Phys.\ Rev.\ B {\bf #2}, #3 (#4)} 
\def \prl#1#2#3#4{ #1, Phys.\ Rev.\ Lett. {\bf #2}, #3 (#4)} 
\def \nat#1#2#3#4{ #1, Nature {\bf #2}, #3 (#4)} 
\def \sci#1#2#3#4{ #1, Science {\bf #2},#3 (#4)}

\bibitem{Wil13} M.\ M.\ Wilde, {\em Quantum Information Theory},
  Cambridge University Press, 2013 [arXiv:1106.1445]; and references therein.

\bibitem{gate-fidelity} Average and minimum input-output fidelity have been investigated 
as average and minimum {\em gate fidelity} in: 
\pra{M.\ Horodecki, P.\ Horodecki, and R.\ Horodecki}{60}{1888}{1999};
M.\ Nielsen, Phys.\ Lett. {\bf A 303}, 249 (2002);
M.\ D.\ Bowdrey, D.\ K.\ L.\ Oi, A.\ J.\ Short, K.\ Banaszek, J.\ A.\
Jones, Phys.\ Lett. {\bf A 294}, 258 (2002); 
\pra{A.\ Gilchrist, N.\ K.\ Langford, and M.\ A.\ Nielsen}{71}{062310}{2005};
J.\ Emerson, R.\ Alicki, K.\ Zyczkowski, J.\ Opt.\ B: Quantum
Semiclass.\ Opt.\ {\bf 7}, 347 (2005);
N.\ Johnston and D.\ W.\ Kribs, J.\ Phys.\ A: Math.\ Theor.\ {\bf 44},
495303 (2011).

\bibitem{Kin02} C. King, J.\ Math.\ Phys.\ {\bf 43}, 4641 (2002). 

\bibitem{multiplicativity} Immediate by definition: $\tilde F(\cal
  N\tenpow m) = \lim_n F^{(n)}(\cal N\tenpow m) = \left(\lim_n F(\cal
  N\tenpow{m n} )^{1/nm}\right)^m = \tilde F(\cal N)^m $.

\bibitem{WH02} R.\ F.\ Werner, A.\ S.\ Holevo, J.\ Math.\ Phys.\ {\bf
  43}, 4353 (2002).

\bibitem{BB88} J.\ Barzilai, J.\ M.\ Borwein, IMA J.\ Numer.\ Anal.,
  {\bf 8}, 141 (1988).

\end{thebibliography}
\end{document}